\pdfoutput=1

\documentclass[final]{aipproc}

\layoutstyle{6x9}

\received{x}
\accepted{x}

\usepackage{ttct0001}

\usepackage{natbib}

\usepackage{shortvrb}
\MakeShortVerb\|

\makeatletter
   \def\@oddfoot{\reset@font
                 \copyright{} 2004 AIP
                 \hfil\@title
                 \hfil\@date\hfil\thepage}
\makeatother

\begin{document}

\author{A. Camero-Arranz}{
  address={National Space Science and Technology Center, Huntsville, USA},
  email={ascension.camero-arranz@nasa.com},
}
\author{M.H. Finger}{
  address={Universities Space Research Association, Huntsville, USA},
  email={},
}

\author{C.A. Wilson--Hodge}{
  address={NASA/Marshall Space Flight Center, Huntsville, USA},
  email={},
}
\author{P. Jenke}{ 
  address={NASA/Marshall Space Flight Center, Huntsville, USA},
  email={},
}
\author{M.J. Coe}{
  address={University of  Southampton, UK},
  email={},
}
\author{I. Steele}{
  address={Liverpool J. Moore's University, UK},
  email={},
}
\author{I. Caballero}{
  address={AIM-CEA Saclay, Paris, France},
  email={},
}
\author{J. Gutierrez-Soto}{
  address={Instituto de Astrof\'{i}sica de Andaluc\'{i}a, Granada, Spain},
  email={},
}
\author{P. Kretschmar}{
  address={ESA/ESAC, Madrid, Spain},
  email={},
}
\author{J. Suso}{
  address={University of Valencia, Spain},
  email={},
}
\author{V.A.McBride}{
  address={University of  Southampton, UK},
  email={},
}
\author{J. Rodr\'iguez}{
  address={AIM-CEA Saclay, Paris, France},
  email={},
}

\title{A\,0535+26: an X--ray/Optical Tour}
\date{2010/09/07}

\keywords{X--rays, Optical, Pulsars}
\classification{95}
\vspace*{-1.35cm}
\begin{abstract}
 
  We compiled  X--ray and Optical observations of the accreting X-ray binary sytem A\,0535+26 since its discovery in 1975, that will allow us to shed light on the unpredictible behavior of this binary system.   We present the data in terms of the Be-disc interaction with the neutron star companion. In addition, we show recent results from the continous monitoring of this source by the Gamma-ray Burst Monitor (GBM),  on board the Fermi observatory,  since its launch in 2008 June 11.
  
\end{abstract}

\maketitle


\vspace*{-1.1cm}
\section{INTRODUCTION}

\begin{figure}[!t]
 \includegraphics[width=8cm,height=6.7cm]{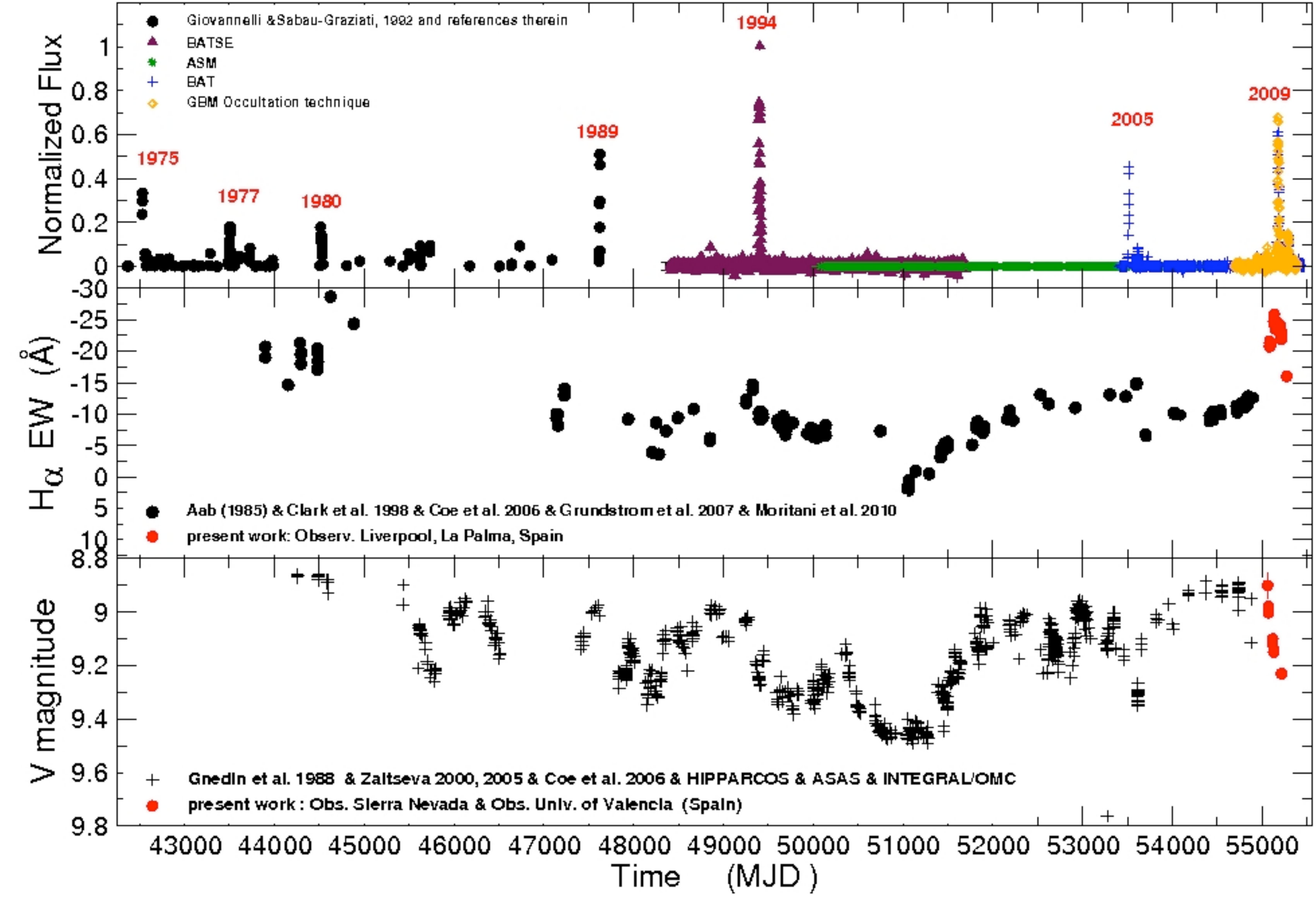}
 \hspace*{0.1cm}
 \includegraphics[width=7.8cm,height=6.9cm]{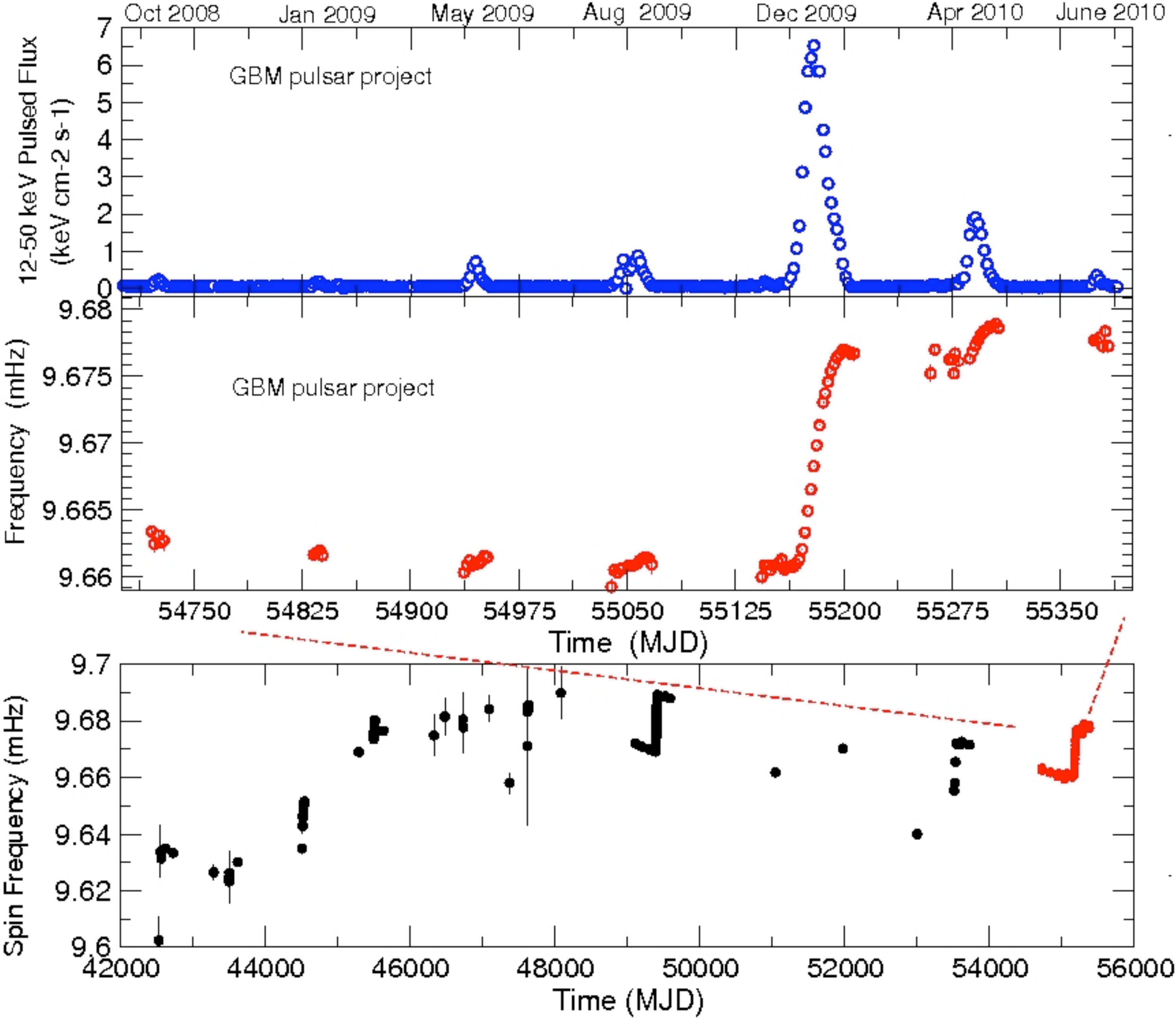}
 \caption{Left. (Top) Long-term X-ray flux history of A\,0535+26. The flux measurements were converted in to Crab units and then normalized using the maximum value for the whole data set (BATSE). (Middle)\,  Long-term H$_\alpha$ history  and photometric study of HDE\,245770 (bottom). Right. (Top)  12-50\,keV pulsed flux since 2008. (Middle) Zoom of the frequency history of this source since 2008 by GBM. (Bottom) Long-term frequency history of A\,0535+26 since 1975. }
\label{long_term}
\end{figure}

A\,0535+26 is a transient Be/X-ray binary pulsar, discovered by Ariel\,V  in 1975 (\cite{rosenberg75}, \cite{coe75}). The orbit is eccentric (e$\sim$0.47) (\cite[and references therein]{finger94}) and the orbital period is 111.10(3) days  with a pulse period  of $\sim$103\,s \cite{finger96}. The optical counterpart is the O9.7\,IIe star HDE\,245770 \citep{liller&eachus75}.  Broad Quasi Periodic Oscillations (QPO) from 27 to 72 mHz were detected  in 1996 by \cite{finger96}, confirming the presence of an accretion disk.  During the 1994 giant outburst, two cyclotron  resonance scattering features at 45 keV  and 100 keV inferred a magnetic field of B$\sim$4$\times$10$^{12}$\,G  \citep{kendziorra94}.   A very nice review on  this binary system by \cite{giovannelli&graziati92} collects X-ray/Optical observations by different missions from 1970 until 1989. Figure~\ref{long_term} shows a long-term X--ray/Optical overview of A\,0535+26/HDE\,245770 from 1975 to 2010.  After the giant outbursts in 1975 (the discovery), 1977, 1980 and 1989,  it was BATSE  which detected this source back in 1994 \citep{bildsten97}.   Three normal outbursts preceded  the 1994 giant one and  two small outbursts took place after that one. Then  A\,0535+26 went back to quiescence for almost 11 years.  A\,0535+26 renewed activity in May 2005 with another giant outburst and  two normal outbursts, becoming a quiet source up to 2008. Figure~\ref{long_term} (middle-left panel) shows a long-term H$_\alpha$ history for HDE\,245770.  The data sets span the dates from  Jan 1979 to Jan 2009 ( \cite[and references therein]{moritani10}).  The bottom-left panel of the same Fig. shows a long-term photometric study  from 1981 to 2005 (\cite{gnedin88}, \cite{lyuty_zaitseva00}, \cite{zaitseva05} and \cite{coe06}).  In this work, we present an Optical/X--ray study of the Be/X-ray binary system A\,0535+26 in an attempt to characterize the  overall picture.

\vspace*{-2cm}
\section{OBSERVATIONS AND ANALYSIS}
\vspace*{-0.2cm}

Our main  X-ray observations in this work come from \textit{Fermi}/GBM.  
The  GBM is an all-sky instrument sensitive to X--rays and gamma rays with energies
between $\sim$8\,keV and $\sim$40\,MeV \citep{gbmpaper}.  GBM includes 12 Sodium Iodide (NaI; 8keV-1MeV)  and 2 Bismuth Germanate (BGO; 150\,keV-40\,MeV) scintillation detectors.  Timing analysis was carried out with CTIME data (8 channel spectra every 0.256\,s)  from only the NaI detectors. 
 \vspace*{-0.02cm}
 
A detailed explanation of our timing technique can be found in \cite{camero2010}.   The 10-15 keV total flux was obtained using the Earth Occultation technique with CSPEC data (128 energy channels every 4.096\,s)  (\cite{colleen09}) . In the present work we also include photometric archival observations from the HIPPARCOS main catalog, the ASAS catalog, \textit{INTEGRAL}/OMC, and  observations 
from the Spanish Astronomical Observatories of Sierra Nevada (OASN) and  University of Valencia (OAUV) .  We transformed the magnitudes from the HIPPARCOS filter to Visual magnitude,  following the procedure found in \cite{harmanec98} and applying B--V and U--B values from \cite{lyuty_zaitseva00}. In addition, we extracted the V magnitude from \cite{coe06}  based on the work done by \cite{grundstrom07}.   The spectroscopic observations come from the Liverpool Telescope (La Palma Observatory, Spain).

\begin{figure}[!]
 \includegraphics[width=5.3cm,height=4.45cm]{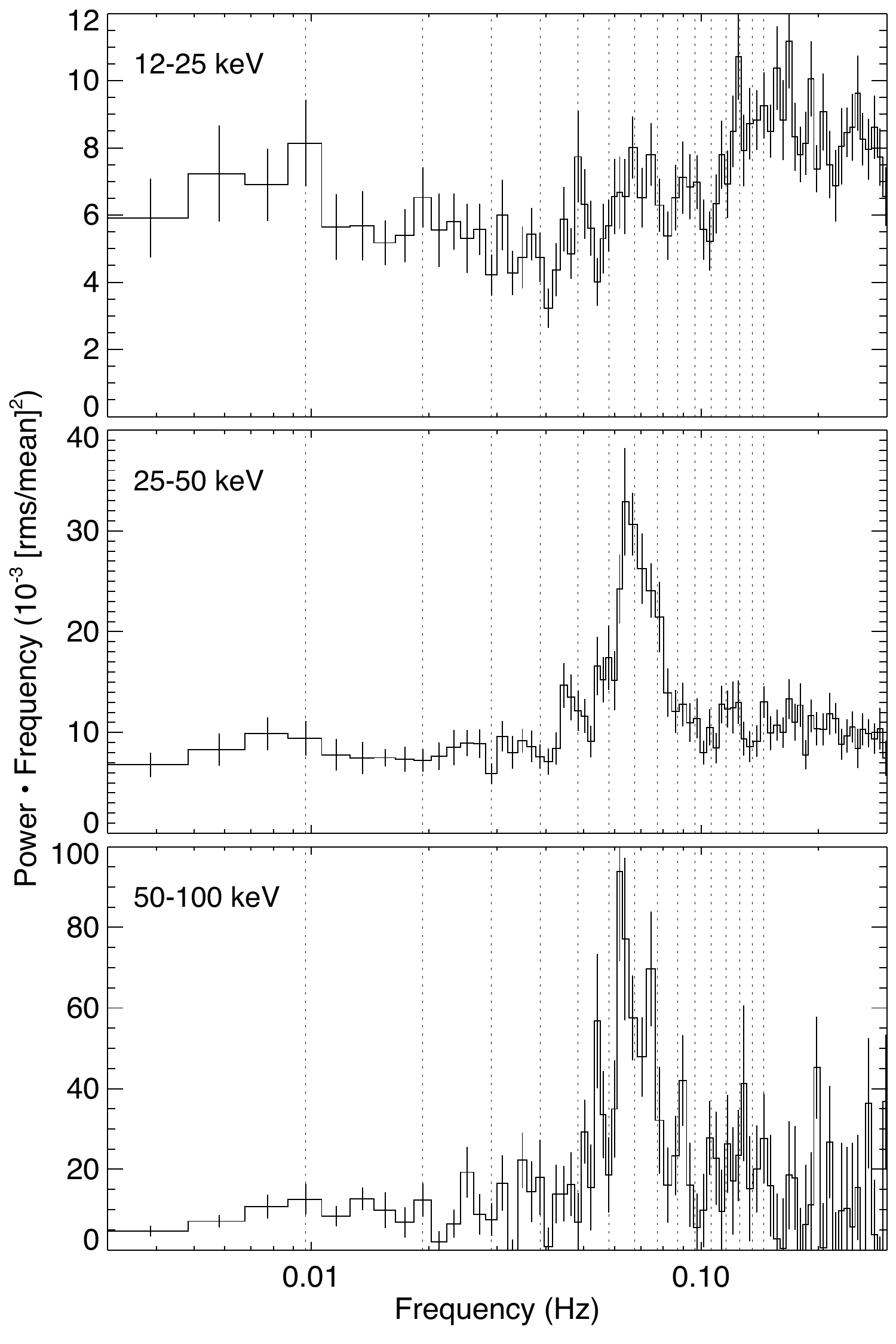}
 \includegraphics[width=4.9cm,height=5.1cm]{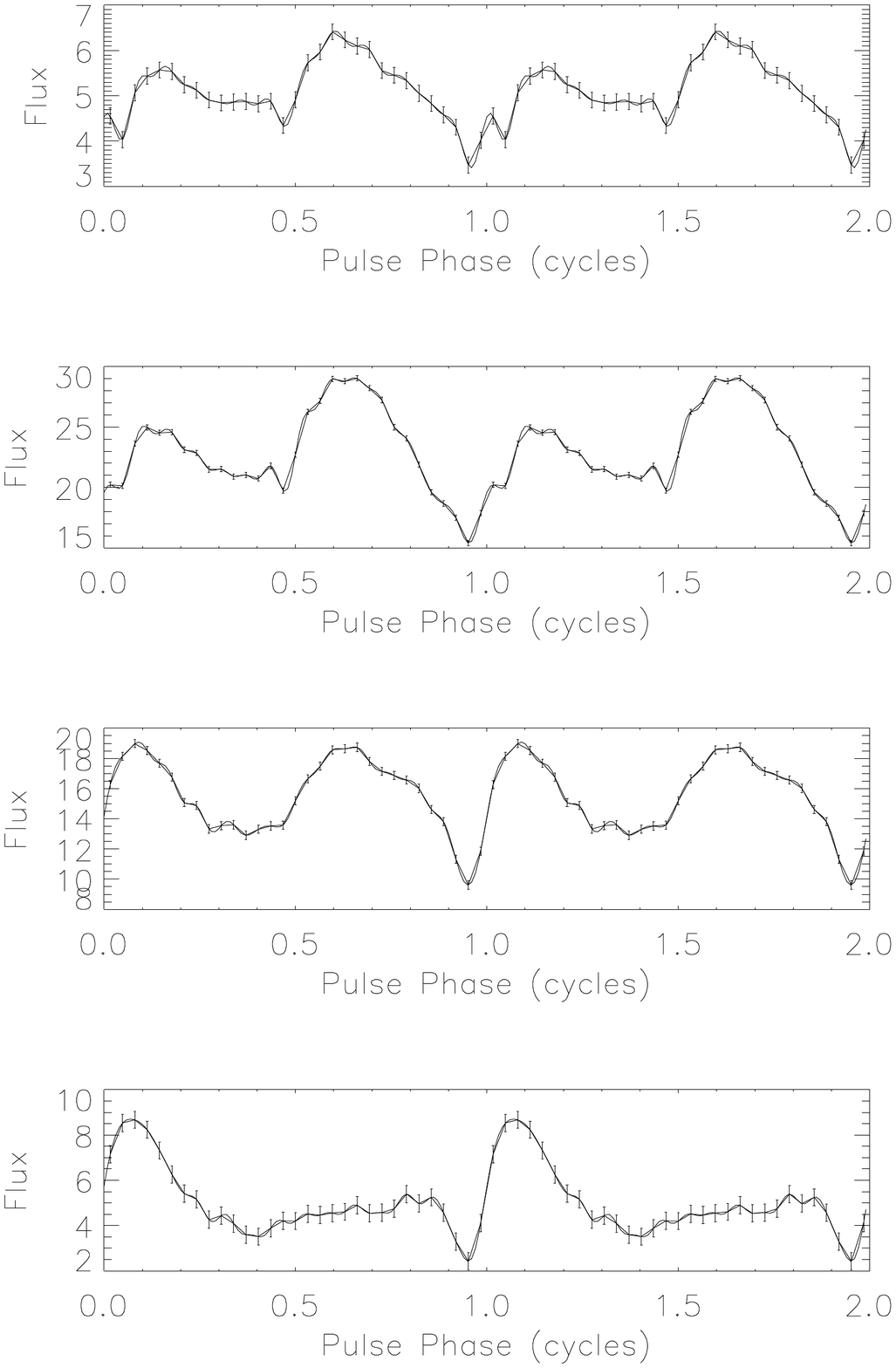}
\hspace*{-0.9cm}
  \includegraphics[width=6.45cm,height=4.7cm]{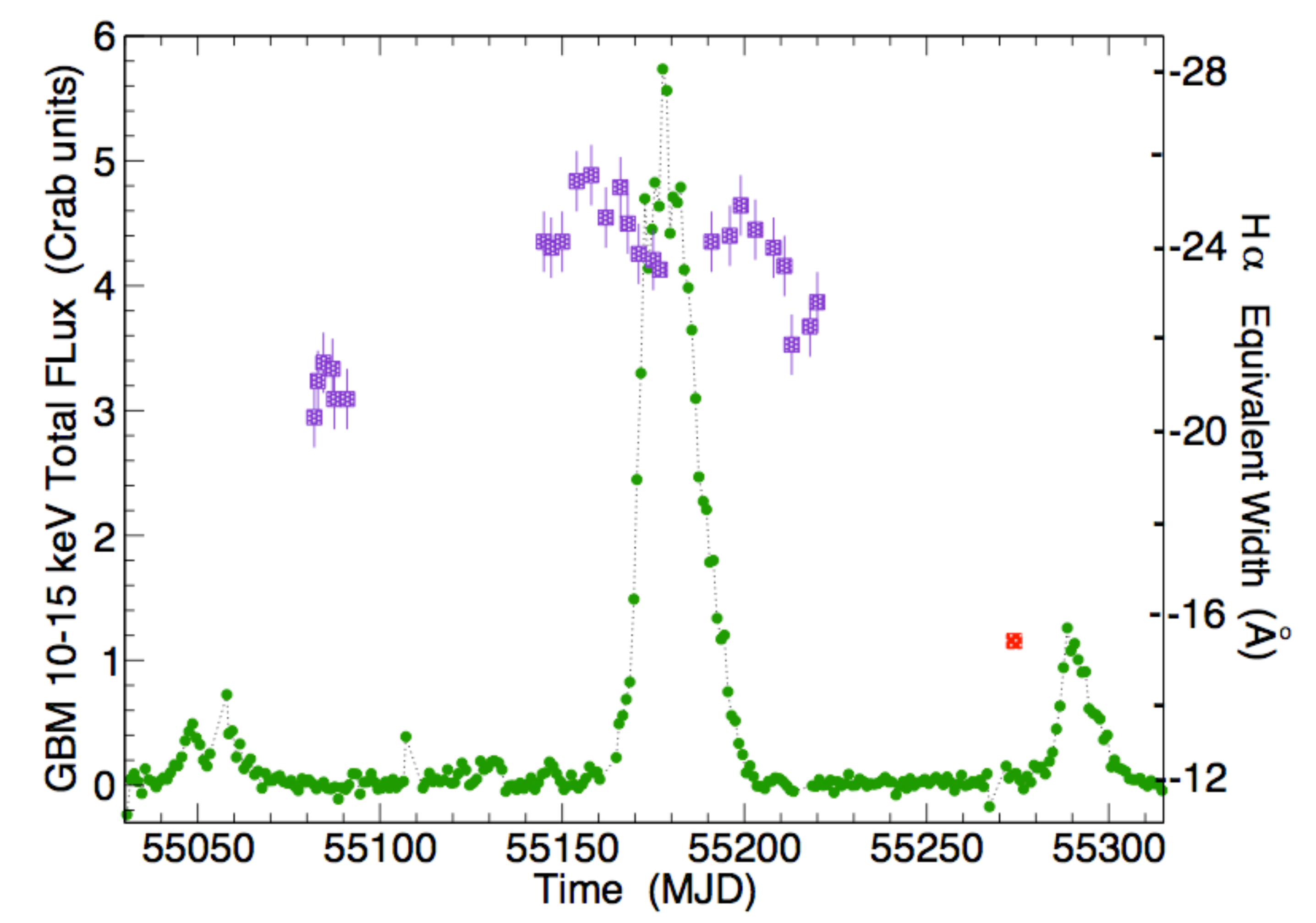}
\caption{Left. QPO centered at 62 mHz from a GBM observation on Dec 11  in three energy bands. Middle. Pulse profile evolution  during 2009 giant outburst near the maximun (from top to bottom: 8-12,\,12-25,\,25-50,\,50-100\,keV). Right. Composite H$_\alpha$-Equivalent-Width/X-ray flux plot. The last red square is from \cite{giovannelliATel2010}.\label{qpo}}
\end{figure}
 
\vspace*{-0.65cm}
\section{RESULTS}

\vspace*{-0.3cm}
\subsubsection*{Pulse Frequency History and Quasi Periodic Oscillations}
\vspace*{-0.05cm}

Fig.~\ref{long_term}\,(right)  shows the pulse frequency history and pulsed flux obtained for A\,0535+26  (more information in http://gammaray.nsstc.nasa.gov/gbm/science/pulsars/).  The  correlation between the X-ray flux and the spin-up rate confirms the presence of an accretion disk in the neutron star.  QPO features correlated with flux/spin-up are further indication of an accretion disk. We analyzed  the aperiodic variability in the X-ray flux of A\,0535+26 following the method described in \cite{finger96}. During the Dec 2009 giant outburst we found that the power spectra of the hard--X ray flux of this source,  between 3 mHz and 1 Hz, consisted of an approximately 1/f power law continuum plus a broad QPO  and a pulse component \citep{fingerwilsoncamero2009}.  On Dec 10 the QPO was centered on 62$\pm$1 mHz (FWHM of 29$\pm$2 mHz) in the 25-50 keV band,  and we were able to detect it from Dec 4  to 27.    The  QPO was strong in the 50--100 keV band (see Fig.~\ref{qpo}, left panel), but not detected in the 12--25 keV range (Finger et al. 2010, in preparation). 

\vspace*{-0.71cm}
\subsubsection*{Pulse Profiles}
\vspace*{-0.15cm}

The  Aug 2009 normal outburst has a double peak structure,  similar to that observed  in the normal outburst right after the 2005 giant one (\cite{finger06}, \cite{caballero08}).  For the 2005 normal outburst, we found that the pulse profiles in the first spike were different than in the main peak of the outburst, suggesting that a different physical mechanism operates in the two peaks.
However, for the  double-peaked normal outburst in Aug 2009, we did not find any difference in pulse shape between the two peaks. For the first time a low energy pulse profile of A\,0535+26 during a giant outburst has been obtained. Fig.~\ref{qpo} (middle) shows the pulse profile evolution with energy for an observation near the peak time of the outburst. We can see how the two main components of the profile  evolve from a weak first peak plus a second strong component, to a strong first peak plus a severely diminished second one (Camero-Arranz et al. 2010, in preparation). 

\vspace*{-0.71cm}
\subsubsection*{X--ray/Optical Correlations}
\vspace*{-0.15cm}

Among other X-ray missions, GBM detected four normal outbursts in 2008 with higher flux outbursts beginning at an earlier orbital phase.  After the Dec 2009 giant outburst, a very large event was predicted based on new X-ray activity detected 19 days before periastron.  Our simultaneous optical observations (Fig.~\ref{qpo}, right)  showed how the circumstellar disk has grown to its full size before the Dec 2009 giant outburst, waiting for the neutron star to arrive and begin accretion. They also revealed a strong H$_{\alpha}$ line in emission during and after the Dec 2009 giant outburst, indicating that the donor star presented a disk around its equator before and after the large event,  suggestive of a residual accretion disk present around the neutron star.  Later on,  strong optical activity of HDE\,245770 was observed on March 19, 2010 by \cite{giovannelliATel2010},  an indicator of an incoming X-ray outburst of the system.  This outburst peaked above $\sim$ 1 Crab  (another giant one?).  
 
\vspace*{-0.5cm}
\section{SUMMARY AND CONCLUSIONS}
\vspace*{-0.2cm}

As we have seen, clues to the unpredictable X-ray behavior of A0535+26 are unveiled with observations of the strong Optical/X-ray correlation observed.  The overall picture is a system going through periods when the donor star has minimal circumstellar disc and then a dramatic disk recovery leads to a flare of X-ray emission. The Be circumstelar disk was exceptionally large just before the Dec 2009 giant outburst,  most likely the origin of the unusual X-ray activity. In particular, this might explain the other large outburst occurring after the Dec 2009 Giant outburst.  The double peak structure of the normal outbursts in 2005 and 2009 remains a mystery.  We obtained the first pulse profile evolution during a giant outburst with GBM, showing the expected double peak structure  evolving with the peak's strength shifting around the 25-50 keV band. Another surprise was the lack of detection of the the X-ray QPO below 25 keV.  Work is ongoing to study the data from different X-ray  missions to determine the timing and spectral properties of this source during the Dec 2009 giant outburst and afterwards. Preliminary results show no significant variation of the cyclotron line energy with the luminosity of the system  (Caballero et al. 2010, in preparation).

\textbf{Acknowlegments}.  We thanks all the {\it Fermi}/GBM team for its help, and  especially the GBM Occultation and Pulsar teams.

\end{document}